\documentstyle[prd,aps,preprint,tighten,epsfig]{revtex}

\begin{document}

\draft

\title{Impact of Fermion Mass Degeneracy on Flavor Mixing}
\author{{\bf Jian-wei Mei} ~ and ~ {\bf Zhi-zhong Xing}}
\address{CCAST (World Laboratory), P.O. Box 8730, Beijing 100080, China \\
and Institute of High Energy Physics, 
Chinese Academy of Sciences, \\
P.O. Box 918 (4), Beijing 100039, China 
\footnote{Mailing address} \\
({\it Electronic address: jwmei@mail.ihep.ac.cn;
xingzz@mail.ihep.ac.cn}) }
\maketitle

\begin{abstract}
We carry out a systematic analysis of flavor mixing and CP violation
in the conceptually interesting limit where two quarks or leptons of
the same charge are degenerate in mass. We pay some particular attention 
to the impact of neutrino mass degeneracy and Majorana phase degeneracy
on the lepton flavor mixing matrix.
\end{abstract}


\newpage

\section{Introduction}

The phenomenon of quark flavor mixing, described by the $3\times 3$
Cabibbo-Kobayashi-Maskawa (CKM) matrix $V^{~}_{\rm CKM}$ \cite{CKM}, 
is attributed to the mismatch between the diagonalization of the 
up-type quark mass matrix $M_{\rm u}$ and that of the down-type
quark mass matrix $M_{\rm d}$. In the standard model, the weak 
charged current of quarks can be written as
\begin{equation}
J^\mu_{\rm quark} \; =\; \overline{(u, c, t)^{~}_{\rm L}} ~ 
\gamma^\mu V^{~}_{\rm CKM} \left ( \matrix{
d \cr
s \cr
b \cr} \right )_{\rm L} \; ,
\end{equation}
where $(u, c, t)$ and $(d, s, b)$ denote the mass eigenstates of
up-type and down-type quarks, respectively. 
The phases of six left-handed quark fields in $J^\mu_{\rm quark}$
can arbitrarily be redefined, but the mass terms
\begin{equation}
\overline{(u, c, t)^{~}_{\rm L}} ~ \left ( \matrix{
m_u & 0 & 0 \cr
0 & m_c & 0 \cr
0 & 0 & m_t \cr} \right ) \left ( \matrix{
u \cr
c \cr
t \cr} \right )_{\rm R} 
\end{equation}
and
\begin{equation}
\overline{(d, s, b)^{~}_{\rm L}} ~ \left ( \matrix{
m_d & 0 & 0 \cr
0 & m_s & 0 \cr
0 & 0 & m_b \cr} \right ) \left ( \matrix{
d \cr
s \cr
b \cr} \right )_{\rm R} 
\end{equation}
do not change if six right-handed quark fields are accordingly 
rephased. This freedom, together with the requirement that
$V^{~}_{\rm CKM}$ must be unitary, allows us to parametrize $V$ 
in terms of four independent parameters, which are commonly
taken as three rotation angles and one CP-violating phase \cite{FX00}.
All of these four parameters have been measured in a number of
delicate experiments on quark mixing and CP violation \cite{PDG}.
 
Thanks to the Super-Kamiokande \cite{SK}, SNO \cite{SNO}, 
KamLAND \cite{KM} and K2K \cite{K2K} neutrino oscillation experiments,
we are now convinced that neutrinos are also massive 
and lepton flavors are mixed. The lepton flavor mixing matrix 
$V^{~}_{\rm MNS}$, which is referred to as the Maki-Nakagawa-Sakata 
(MNS) matrix \cite{MNS} in some literature,
arises from the mismatch between the diagonalization of the charged
lepton mass matrix $M_l$ and that of the (effective) neutrino mass 
matrix $M_\nu$ at low energies. Similar to $J^\mu_{\rm quark}$, the
weak charged current of leptons reads 
\begin{equation}
J^\mu_{\rm lepton} \; =\; \overline{(e, \mu, \tau)^{~}_{\rm L}} ~ 
\gamma^\mu V^{~}_{\rm MNS} \left ( \matrix{
\nu^{~}_1 \cr
\nu^{~}_2 \cr
\nu^{~}_3 \cr} \right )_{\rm L} \; ,
\end{equation}
where $(e, \mu, \tau)$ and $(\nu^{~}_1, \nu^{~}_2, \nu^{~}_3)$ denote 
the mass eigenstates of charged leptons and active neutrinos, 
respectively. If neutrinos are Dirac particles, the MNS matrix 
$V^{~}_{\rm MNS}$ is just a leptonic analogue of the CKM matrix 
$V^{~}_{\rm CKM}$ and can also be parametrized in terms of three mixing 
angles and one CP-violating phase. Only the two mixing angles
relevant to solar and atmospheric neutrino oscillations have so far
been determined from a global analysis of current experimental 
data \cite{Fit}. If neutrinos are Majorana particles, however, 
there is no freedom to redefine the {\it relative} phases of three
left-handed neutrino fields in $J^\mu_{\rm lepton}$. The reason is
simply that the effective Majorana mass term
\begin{equation}
\overline{(\nu^{~}_1, \nu^{~}_2, \nu^{~}_3)^{~}_{\rm L}} ~ 
\left ( \matrix{
m^{~}_1 & 0 & 0 \cr
0 & m^{~}_2 & 0 \cr
0 & 0 & m^{~}_3 \cr} \right ) \left ( \matrix{
\nu^{\rm c}_1 \cr
\nu^{\rm c}_2 \cr
\nu^{\rm c}_3 \cr} \right )_{\rm R} 
\end{equation}
depends nontrivially on the rephasing of left-handed neutrino 
fields (where $\nu^{\rm c}_i \equiv C \overline{\nu}^T_i$ with
$C$ being the charge-conjugation operator). In this case,
two additional CP-violating phases are needed to fully parametrize 
$V^{~}_{\rm MNS}$ -- namely, a complete parametrization of
$V^{~}_{\rm MNS}$ requires three rotation angles and three 
CP-violating phases \cite{FX01}. Because neutrinos are 
expected to be of the Majorana type in most of the promising neutrino 
mass models (such as those incorporating the seesaw mechanism \cite{SS}),
we shall focus our attention on the phenomenon of lepton flavor mixing 
associated with three light Majorana neutrinos in the following.

The parametrization of $V^{~}_{\rm CKM}$ or $V^{~}_{\rm MNS}$ can
concretely be realized with the help of three orthogonal matrices
$O_{12}$, $O_{23}$ and $O_{13}$, which correspond to simple rotations 
in the (1,2), (2,3) and (1,3) planes: 
\begin{eqnarray}
O_{12}(\theta_{12}) & = &
\left ( \matrix{
c_{12} & s_{12} & 0 \cr
-s_{12} & c_{12} & 0 \cr
0 & 0 & 1 \cr} \right ) \; ,
\nonumber \\ 
O_{23}(\theta_{23}) & = &
\left ( \matrix{
1 & 0 & 0 \cr
0 & c_{23} & s_{23} \cr
0 & -s_{23} & c_{23} \cr} \right ) \; ,
\nonumber \\ 
O_{13}(\theta_{13}) & = &
\left ( \matrix{
c_{13} & 0 & s_{13} \cr
0 & 1 & 0 \cr 
-s_{13} & 0 & c_{13} \cr} \right ) \; ,
\end{eqnarray}
where $s_{ij} \equiv \sin\theta_{ij}$ and $c_{ij} \equiv \cos\theta_{ij}$.
Of course, at least one complex phase should properly be included 
into $O_{ij}$ (e.g., by replacing ``1'' in the (3,3) position of $O_{12}$ 
with $e^{i\delta}$ \cite{FX98}), such that $V^{~}_{\rm CKM}$ consists
of one nontrivial CP-violating phase and $V^{~}_{\rm MNS}$ contains
three nontrivial CP-violating phases. For the sake of convenience in
subsequent discussions, we list twelve different representations of 
$V^{~}_{\rm CKM}$ and $V^{~}_{\rm MNS}$ in Table I.

As pointed out by Jarlskog \cite{J}, it is always possible to make one 
matrix element of $V^{~}_{\rm CKM}$ vanishing in the limit where
two quarks of the same charge are degenerate in mass. In
such an unrealistic but conceptually interesting case, the 
nontrivial CP-violating phase is also removable from $V^{~}_{\rm CKM}$. 
One may further speculate
(a) how the quark mass degeneracy explicitly affects $V^{~}_{\rm CKM}$;
(b) whether similar results can be obtained for $V^{~}_{\rm MNS}$ in 
the limit where two charged leptons or two neutrinos are degenerate in 
mass; and (c) what impact the degeneracy of two Majorana CP-violating 
phases together with the degeneracy of two neutrino masses may have on 
$V^{~}_{\rm MNS}$. Such speculations call for a careful analysis of the 
underlying correlation between fermion mass degeneracy and flavor mixing.

The purpose of this paper is to present a systematic analysis of the
impact of fermion mass degeneracy on flavor mixing and CP violation, 
in order to answer
the above questions. Section II is devoted to quark flavor mixing in 
the limit of quark mass degeneracy. Detailed discussions about lepton 
flavor mixing in the limit of lepton mass degeneracy are given in
Section III. We draw the conclusion in Section IV. 

\section{Quark flavor mixing}

We first discuss how the CKM matrix $V^{~}_{\rm CKM}$ can be simplified
in the limit where two quarks of the same charge are degenerate in mass.
To do so, it is convenient to adopt one of the parametrizations of 
$V^{~}_{\rm CKM}$ listed in Table I.
Let us assume $u$ and $c$ quarks to be degenerate as an example. In
this case, it would be impossible to distinguish between $u$ and $c$
-- in other words, there would be an extra symmetry which allows an
arbitrary {\it unitary} rotation in the space spanned by $u$ and $c$
quarks but keeps the mass term in Eq. (2) unchanged \cite{FX99}. Making
the transformation
\begin{equation}
\left ( \matrix{
u \cr
c \cr
t \cr} \right ) 
\; \Longrightarrow \; 
\left ( \matrix{
u \cr
c \cr
t \cr} \right )' = \; O^\dagger_{12}
\left ( \matrix{
u \cr
c \cr
t \cr} \right ) \; ,
\end{equation} 
one may easily verify that the primed fields remain the mass 
eigenstates of three up-type quarks. In the new basis, the CKM matrix 
becomes $V'_{\rm CKM} = O^\dagger_{12} V^{~}_{\rm CKM}$. Once 
parametrizations (A), (B), (G) and (H) in Table I are taken into 
account, we obtain the corresponding patterns of $V'_{\rm CKM}$
as follows:
\begin{equation}
V'_{\rm CKM} \; = \; \left \{ \matrix{
O_{23} \otimes P_1 \otimes \tilde{O}_{12} \;\;\;\; {\rm (A)} \; , \cr
O_{13} \otimes P_2 \otimes \tilde{O}_{12} \;\;\;\; {\rm (B)} \; , \cr
O_{23} \otimes P_1 \otimes O_{13} \;\;\;\; {\rm (G)} \; , \cr
O_{13} \otimes P_2 \otimes O_{23} \;\;\;\; {\rm (H)} \; , \cr} 
\right .
\end{equation}
in which $(V'_{\rm CKM})_{13} =0$, $(V'_{\rm CKM})_{23} =0$,
$(V'_{\rm CKM})_{12} =0$ and $(V'_{\rm CKM})_{21} =0$,
respectively.
This result indicates that one of the four off-diagonal matrix 
elements in the first two rows of $V^{~}_{\rm CKM}$
(or equivalently, one of the three mixing angles of 
$V^{~}_{\rm CKM}$) is removable in the $m_u = m_c$ limit. Because
$O_{23}$ (or $O_{13}$) and $P_1$ (or $P_2$) are commutable, the
latter can then be rotated away from $V'_{\rm CKM}$ by rephasing
the up-type quark mass eigenstates. 
It becomes obvious that no CP violation can really manifest
itself in the $m_u = m_c$ limit. We conclude that there would 
exist only two flavor mixing angles and the CP symmetry would 
be conserving, if $u$ and $c$ quarks were degenerate in mass.

The above arguments can straightforwardly be repeated for any 
pair of up-type or down-type quarks. To be concise, we summarize 
the relevant results in Table II, where the removable complex 
phase in $V'_{\rm CKM}$ has been omitted. We see that it is always 
possible to make one of the six off-diagonal matrix elements of 
$V^{~}_{\rm CKM}$ vanishing in the limit where two quarks of the 
same charge are degenerate in mass. CP would be a good symmetry 
in this case.

If both $m_u = m_c$ and $m_d = m_s$ held, one could easily
show that the transformation
\begin{equation}
\left ( \matrix{
d \cr
s \cr
b \cr} \right ) \; \Longrightarrow \;
\left ( \matrix{
d \cr
s \cr
b \cr} \right )' = \; \tilde{O}_{12}
\left ( \matrix{
d \cr
s \cr
b \cr} \right ) 
\end{equation}
together with that in Eq. (7) does not change any physics.
In the new (primed) basis, the CKM matrix $V^{~}_{\rm CKM}$
turns out to be 
$V''_{\rm CKM} = O^\dagger_{12} V^{~}_{\rm CKM} 
\tilde{O}^\dagger_{12}$. It is then possible to obtain
\begin{equation}
~~~~ V''_{\rm CKM} \; = \; \left \{ \matrix{
O_{23} \otimes P_1 \;\;\;\;\;\;\; {\rm (A)} \; , \cr
O_{13} \otimes P_2 \;\;\;\;\;\;\; {\rm (B)} \; , \cr}
\right .
\end{equation}
for parametrizations (A) and (B) of $V^{~}_{\rm CKM}$. The phase 
matrix $P_1$ or $P_2$ on the right-hand side of $V''_{\rm CKM}$ 
can easily be rotated away by redefining the relevant phases of 
down-type quark fields. Eq. (10) indicates that only one of the 
three mixing angles of $V^{~}_{\rm CKM}$ can survive in the limit 
of $m_u = m_c$ and $m_d = m_s$. The similar argument is applicable
for any pair of up-type quarks and any pair of down-type quarks.
There are totally nine such possibilities, as summarized in
Table III, where the removable complex phase in $V''_{\rm CKM}$ 
has been omitted.

If three up-type (or down-type) quarks were all degenerate in mass,
one would have sufficient freedom to redefine the quark mass
eigenstates and transform the CKM matrix $V^{~}_{\rm CKM}$ into
the unity matrix. Then there would be no quark flavor mixing and CP
violation. We conclude that a necessary condition for three-flavor 
mixing and CP violation in the quark sector is that two quarks 
of the same charge must not be degenerate in mass. This condition
is certainly satisfied in reality, because three quarks in each 
sector have a strong mass hierarchy \cite{PDG}.

\section{Lepton flavor mixing}

The Majorana nature of $(\nu^{~}_1, \nu^{~}_2, \nu^{~}_3)$ neutrinos 
makes the description of lepton flavor mixing somehow more
complicated than the description of quark flavor mixing. To 
make our discussions more convenient, we decompose 
$V^{~}_{\rm MNS}$ into a product of $U^{~}_{\rm MNS}$ and $P_\phi$; 
i.e., $V^{~}_{\rm MNS} = U^{~}_{\rm MNS} P_\phi$, where $U^{~}_{\rm MNS}$ 
is just a leptonic analogue of the CKM matirx $V^{~}_{\rm CKM}$ and 
$P_\phi = {\rm Diag}\{e^{i\phi_1}, e^{i\phi_2}, e^{i\phi_3}\}$ 
represents the diagonal Majorana phase matrix containing two nontrivial 
CP-violating phases \cite{FX01}. Subsequently we take two steps to 
discuss the impact of lepton mass degeneracy on lepton flavor mixing.

First, we consider the partial or total mass degeneracy of three charged
leptons, whose left-handed and right-handed fields may have completely
independent phases. If $m_e = m_\mu$ held, for example, one would 
be able to make one of the four off-diagonal matrix elements in the
first two rows of $U^{~}_{\rm MNS}$ vanishing. This situation is 
exactly the same as the situation of quark mixing in the $m_u = m_c$ 
limit. Thus the relevant discussions in Section II are applicable
for $U^{~}_{\rm MNS}$, as shown in Table IV, where the reduced form of 
$U^{~}_{\rm MNS}$ has been denoted by $U'_{\rm MNS}$. Note, however, 
that the Majorana phase matrix $P_\phi$ is not influenced by the mass
degeneracy of charged leptons. 

Second, we consider the case in which three neutrinos are partially or 
totally degenerate in mass. If $m_1 = m_2$ held, for instance, there 
would be an extra symmetry which allows an arbitrary {\it orthogonal}
rotation in the space spanned by $\nu_1$ and $\nu_2$ neutrinos but
keeps the mass term in Eq. (5) unchanged. Namely, the transformation
\begin{equation}
\left ( \matrix{
\nu^{~}_1 \cr
\nu^{~}_2 \cr
\nu^{~}_3 \cr} \right ) 
\; \Longrightarrow \; 
\left ( \matrix{
\nu^{~}_1 \cr
\nu^{~}_2 \cr
\nu^{~}_3 \cr} \right )' = \; \tilde{O}_{12}
\left ( \matrix{
\nu^{~}_1 \cr
\nu^{~}_2 \cr
\nu^{~}_3 \cr} \right ) \; 
\end{equation} 
does not change any physics. In the new basis, the MNS matrix becomes
$V'_{\rm MNS} = V^{~}_{\rm MNS} \tilde{O}^T_{12}
= U^{~}_{\rm MNS} P_\phi \tilde{O}^T_{12}$. The explicit
parametrizations of $V^{~}_{\rm MNS}$ have been listed in Table I.
Note that $\tilde{O}^T_{12}$ cannot commute with 
$P_\phi$, unless $\phi_1 = \phi_2$ is taken. Hence $V'_{\rm MNS}$ is
{\it in general} unable to be simplified by the orthogonal transformation
in Eq. (11). Such a conclusion can also be drawn for $m_1 = m_3$ and
$m_2 = m_3$ cases. It remains valid even in the $m_1 = m_2 = m_3$ case. 

We proceed with the decomposition 
$V^{~}_{\rm MNS} = U^{~}_{\rm MNS} P_\phi$. The Majorana 
phase matrix $P_\phi$ can be absorbed by a redefinition of the neutrino 
mass eigenstates,
\begin{equation}
\left ( \matrix{
\nu^{~}_1 \cr
\nu^{~}_2 \cr
\nu^{~}_3 \cr} \right ) \Longrightarrow
\left ( \matrix{
\nu^{~}_1 \cr
\nu^{~}_2 \cr
\nu^{~}_3 \cr} \right )' = \; P_\phi 
\left ( \matrix{
\nu^{~}_1 \cr
\nu^{~}_2 \cr
\nu^{~}_3 \cr} \right ) \; .
\end{equation}
In the new basis, the neutrino mass term in Eq. (5) turns out to be
\begin{equation}
\overline{(\nu^{~}_1, \nu^{~}_2, \nu^{~}_3)'_{\rm L}} ~ 
\left ( \matrix{
m'_1 & 0 & 0 \cr
0 & m'_2 & 0 \cr
0 & 0 & m'_3 \cr} \right ) \left ( \matrix{
\nu^{\rm c}_1 \cr
\nu^{\rm c}_2 \cr
\nu^{\rm c}_3 \cr} \right )'_{\rm R} \; , 
\end{equation}
where $m'_a \equiv m_a e^{2i\phi_a}$ (for $a=1,2,3$). The complex
neutrino masses $(m'_1, m'_2, m'_3)$ contain the information about
Majorana phases. In the limit where $m'_1 = m'_2$ holds (i.e., both 
$m_1 = m_2$ and $\phi_1 = \phi_2$ hold
\footnote{Because $P_\phi$ is always associated with $U^{~}_{\rm MNS}$ 
for a given parametrization of $V^{~}_{\rm MNS}$, the assumption of
Majorana phase degeneracy (such as $\phi_1 = \phi_2$) is actually 
dependent on which explicit representation or phase convention has 
been adopted for $U^{~}_{\rm MNS}$.}),
one may transform the neutrino mass eigenstates in a simple way 
similar to Eq. (11),
\begin{equation}
\left ( \matrix{
\nu^{~}_1 \cr
\nu^{~}_2 \cr
\nu^{~}_3 \cr} \right )' \Longrightarrow
\left ( \matrix{
\nu^{~}_1 \cr
\nu^{~}_2 \cr
\nu^{~}_3 \cr} \right )'' = \; 
\tilde{O}_{12} \left ( \matrix{
\nu^{~}_1 \cr
\nu^{~}_2 \cr
\nu^{~}_3 \cr} \right )' \; .
\end{equation}
This leads to $U_{\rm MNS}' = U_{\rm MNS} \tilde{O}^T_{12}$ 
in the new basis. Once parametrizations (A), (B), (J) and (L) in 
Table I are taken into account, we arrive at the corresponding 
patterns of $U'_{\rm MNS}$ as follows:
\begin{equation}
U'_{\rm MNS} \; = \; \left \{ \matrix{
O_{12} \otimes O_{23} \otimes P_1 \;\;\;\; {\rm (A)} \; , \cr
O_{12} \otimes O_{13} \otimes P_2 \;\;\;\; {\rm (B)} \; , \cr
O_{23} \otimes O_{13} \otimes P_2 \;\;\;\; {\rm (J)} \; , \cr
O_{13} \otimes O_{23} \otimes P_1 \;\;\;\; {\rm (L)} \; , \cr} 
\right .
\end{equation}
in which $(U'_{\rm MNS})_{31} =0$, $(U'_{\rm MNS})_{32} =0$,
$(U'_{\rm MNS})_{12} =0$ and $(U'_{\rm MNS})_{21} =0$, respectively.
This result means that one of the four off-diagonal matrix elements 
in the first two columns of $U_{\rm MNS}$ or $V^{~}_{\rm MNS}$
(or equivalently, one of the three mixing angles of 
$V^{~}_{\rm MNS}$) is removable in the $m'_1 = m'_2$ limit. Note
that the complex phase in $U'_{\rm MNS}$, which comes from $P_1$ or
$P_2$, cannot be removed by redefining the mass eigenstates of 
charged leptons and neutrinos. Note also that the neutrino mass
term in Eq. (13) consists of an irremovable Majorana phase in
the $m'_1 = m'_2$ limit (i.e., the phase difference 
$\phi_3 - \phi_1 = \phi_3 - \phi_2$). Thus $V^{~}_{\rm MNS}$ would
totally contain two mixing angles and two CP-violating phases in
the conceptually interesting case where $m'_1 = m'_2$ held. Such
a situation is quite different from the corresponding situation 
of quark flavor mixing.

The above arguments can be repeated for $m'_1 = m'_3$ and
$m'_2 = m'_3$ cases. We summarize the relevant results in Table IV.
It is worth remarking that $V^{~}_{\rm MNS}$ may have one 
nontrivial mixing angle and one nontrivial CP-violating phase even
in the $m'_1 = m'_2 = m'_3$ limit \cite{Branco}. To see this 
point in a more obvious way, let us replace the transformation
in Eq. (14) by
\begin{equation}
\left ( \matrix{
\nu^{~}_1 \cr
\nu^{~}_2 \cr
\nu^{~}_3 \cr} \right )' \Longrightarrow
\left ( \matrix{
\nu^{~}_1 \cr
\nu^{~}_2 \cr
\nu^{~}_3 \cr} \right )'' = \; O_{ij} \tilde{O}_{12} 
\left ( \matrix{
\nu^{~}_1 \cr
\nu^{~}_2 \cr
\nu^{~}_3 \cr} \right )' \; ,
\end{equation}
where $O_{ij} = O_{23}$ for parametrizations (A) and (L)
or $O_{ij} = O_{13}$ for parametrizations (B) and (J). Then we obtain 
$U'_{\rm MNS} = U^{~}_{\rm MNS} \tilde{O}^T_{12} O^T_{ij}$ in
the new basis. Because $O_{23}$ (or $O_{13}$) and $P_1$ (or $P_2$) 
are commutable, $U'_{\rm MNS}$ can be simplified to
\begin{equation}
U'_{\rm MNS} \; = \; \left \{ \matrix{
O_{12} \otimes P_1 \;\;\;\;\;\;\;\;\;\;\;\; {\rm (A)} \; , \cr
O_{12} \otimes P_2 \;\;\;\;\;\;\;\;\;\;\;\; {\rm (B)} \; , \cr
O_{23} \otimes P_2 \;\;\;\;\;\;\;\;\;\;\;\; {\rm (J)} \; , \cr
O_{13} \otimes P_1 \;\;\;\;\;\;\;\;\;\;\;\; {\rm (L)} \; . \cr} 
\right .
\end{equation}
It should be noted that three phases $\phi_i$ can all be removed from 
the neutrino mass term in the $m'_1 = m'_2 = m'_3$ limit (i.e., 
$\phi_1 = \phi_2 = \phi_3$). Hence we are left with one mixing
angle and one CP-violating phase in this special case, as shown
by Eq. (17). 

Finally, we take a look at the case in which a pair of charged 
lepton masses are degenerate and a pair of complex (primed) neutrino 
masses are also degenerate. Let us assume $m_e = m_\mu$ and 
$m'_1 = m'_2$ as an example. Then the orthogonal matrix on the 
left-hand side of $U'_{\rm MNS}$ in Eq. (15) can be rotated away
by a proper redefinition of the mass eigenstates of charged leptons.
Since $O_{23}$ (or $O_{13}$) and $P_1$ (or $P_2$) are commutable,
the latter can also be rotated away by rephasing the charged
lepton fields. We are therefore left with the reduced MNS matrix
$U''_{\rm MNS} = O_{23}$ for parametrizations (A) and (L) or
$U''_{\rm MNS} = O_{13}$ for parametrizations (B) and (J). Taking
account of the nontrivial Majorana phase in the neutrino mass
term (i.e., the phase difference $\phi_3 - \phi_1 = \phi_3 - \phi_2$),
we conclude that the MNS matrix would consist of one mixing angle 
and one CP-violating phase in the limit where both $m_e = m_\mu$ and
$m'_1 = m'_2$ held. The above argument can be repeated for any
pair of charged leptons and any pair of neutrinos, and the relevant
results are summarized in Table V.

Although there is little direct information about the absolute masses
of three neutrinos, the observed phenomena of solar and atmospheric
neutrino oscillations {\it do} forbit any pair of neutrino masses to 
be exactly degenerate. In addition, the strong mass hierarchy of three
charged leptons has long been observed \cite{PDG}. Thus the above 
discussions mainly serve for the conceptual clarification of lepton 
flavor mixing in the limit of lepton mass degeneracy.

\section{Concluding remarks}

We have done a systematic analysis of flavor mixing and CP 
violation in the limit where two quarks or leptons of the same
charge are degenerate in mass. In particular, the impact of
neutrino mass degeneracy and Majorana phase degeneracy on the
lepton flavor mixing matrix has been discussed in some detail.

Although a limit of fermion mass degeneracy is primarily of 
conceptual interest, it might be able to serve as a useful 
starting point of view for building realistic models of flavor 
mixing. For instance, the degeneracy of three neutrino
masses implies a possible S(3) symmetry in the effective 
Majorana neutrino mass matrix \cite{FX96}: the observed neutrino
mass-squared differences may be obtained by breaking that
symmetry in a perturbative way, and a nearly bi-maximal lepton
flavor mixing pattern is also achievable if the charged lepton
mass matrix has an approximate (broken) 
$\rm S(3)_{\rm L} \times S(3)_{\rm R}$ symmetry. 

Understanding the generation of fermion masses and the origin 
of flavor mixing and CP violation has been a big challenge to
particle physicists. Our analysis shows that the phenomenon of
flavor mixing and that of CP violation would become simpler if
three quarks or leptons of the same charge were partially or
totally degenerate in mass. In this sense, possible correlation
between fermion masses and flavor mixing is expected to exist
in a theory more fundamental than the standard model. So is
possible correlation between flavor mixing and CP violation.
We hope that much more experimental progress in flavor
physics could finally help us pin down the dynamics of fermion 
masses, flavor mixing and CP violation.

\acknowledgments{One of us (Z.Z.X.) is grateful to H. Fritzsch and 
X.G. He for helpful communications. This work was supported in part 
by the National Natural Science Foundation of China.}

\newpage

\begin{table}
\caption{Twelve different parametrizations of the CKM matrix
$V^{~}_{\rm CKM}$ and the MNS matrix $V^{~}_{\rm MNS}$. Here
$O_{ij}$ and $\tilde{O}_{ij}$ (for $ij = 12, 13, 23$) both
describe a rotation in the $(i,j)$ plane, but their corresponding
rotation angles are in general different. The phase matrices
$P_i$ (for $i=1,2,3$) and $P_\phi$ are defined as
$P_1 = {\rm Diag}\{e^{i\delta}, 1, 1\}$,
$P_2 = {\rm Diag}\{1, e^{i\delta}, 1\}$, 
$P_3 = {\rm Diag}\{1, 1, e^{i\delta}\}$ and
$P_\phi = {\rm Diag}\{e^{i\phi_1}, e^{i\phi_2}, e^{i\phi_3}\}$,
respectively. Two of the three phases in $P_\phi$ or their
combinations represent the two nontrivial Majorana-type 
CP-violating phases of $V^{~}_{\rm MNS}$.}
\vspace{0.3cm}
\begin{center}
\begin{tabular}{|c|c|c|}
& $V^{~}_{\rm CKM}$ 
& $V^{~}_{\rm MNS}$ \\ \hline
~~ Parametrization (A) ~~ 
& $O_{12} \otimes O_{23} \otimes P_1 \otimes \tilde{O}_{12}$
& $O_{12} \otimes O_{23} \otimes P_1 \otimes \tilde{O}_{12}
\otimes P_\phi$ \\ \hline
Parametrization (B) 
& $O_{12} \otimes O_{13} \otimes P_2 \otimes \tilde{O}_{12}$
& $O_{12} \otimes O_{13} \otimes P_2 \otimes \tilde{O}_{12}
\otimes P_\phi$ \\ \hline
Parametrization (C) 
& $O_{23} \otimes O_{12} \otimes P_3 \otimes \tilde{O}_{23}$
& $O_{23} \otimes O_{12} \otimes P_3 \otimes \tilde{O}_{23}
\otimes P_\phi$ \\ \hline
Parametrization (D) 
& $O_{23} \otimes O_{13} \otimes P_2 \otimes \tilde{O}_{23}$
& $O_{23} \otimes O_{13} \otimes P_2 \otimes \tilde{O}_{23}
\otimes P_\phi$ \\ \hline
Parametrization (E) 
& $O_{13} \otimes O_{12} \otimes P_3 \otimes \tilde{O}_{13}$
& $O_{13} \otimes O_{12} \otimes P_3 \otimes \tilde{O}_{13}
\otimes P_\phi$ \\ \hline
Parametrization (F) 
& $O_{13} \otimes O_{23} \otimes P_1 \otimes \tilde{O}_{13}$
& $O_{13} \otimes O_{23} \otimes P_1 \otimes \tilde{O}_{13}
\otimes P_\phi$ \\ \hline
Parametrization (G) 
& $O_{12} \otimes O_{23} \otimes P_1 \otimes O_{13}$
& $O_{12} \otimes O_{23} \otimes P_1 \otimes O_{13}
\otimes P_\phi$ \\ \hline
Parametrization (H) 
& $O_{12} \otimes O_{13} \otimes P_2 \otimes O_{23}$
& $O_{12} \otimes O_{13} \otimes P_2 \otimes O_{23}
\otimes P_\phi$ \\ \hline
Parametrization (I) 
& $O_{23} \otimes O_{12} \otimes P_3 \otimes O_{13}$
& $O_{23} \otimes O_{12} \otimes P_3 \otimes O_{13}
\otimes P_\phi$ \\ \hline
Parametrization (J) 
& $O_{23} \otimes O_{13} \otimes P_2 \otimes O_{12}$
& $O_{23} \otimes O_{13} \otimes P_2 \otimes O_{12}
\otimes P_\phi$ \\ \hline
Parametrization (K) 
& $O_{13} \otimes O_{12} \otimes P_3 \otimes O_{23}$
& $O_{13} \otimes O_{12} \otimes P_3 \otimes O_{23}
\otimes P_\phi$ \\ \hline
Parametrization (L) 
& $O_{13} \otimes O_{23} \otimes P_1 \otimes O_{12}$
& $O_{13} \otimes O_{23} \otimes P_1 \otimes O_{12}
\otimes P_\phi$ 
\end{tabular}
\end{center}
\end{table}

\newpage

\begin{table}
\caption{The impact of quark mass degeneracy on the CKM matrix. The
symbol ``$\surd$'' means that it is possible to make one off-diagonal
matrix element of $V^{~}_{\rm CKM}$ vanishing in the limit where two 
quarks of the same charge are degenerate in mass.}
\vspace{0.3cm}
\begin{center}
\begin{tabular}{|c|c|c|c|c|c|c|}
& $m_u = m_c$ & $m_u = m_t$ & $m_c = m_t$
& $m_d = m_s$ & $m_d = m_b$ & $m_s = m_b$ \\ \hline
~~ $\matrix{V'_{\rm CKM} = O_{23}\otimes O_{13} \cr
{\rm with} ~ (V'_{\rm CKM})_{12} = 0 \cr}$ ~~
& $\surd$ & $\surd$ & & $\surd$ & & $\surd$ \\ \hline
$\matrix{V'_{\rm CKM} = O_{13}\otimes O_{23} \cr
{\rm with} ~ (V'_{\rm CKM})_{21} = 0 \cr}$
& $\surd$ & & $\surd$ & $\surd$ & $\surd$ & \\ \hline
$\matrix{V'_{\rm CKM} = O_{23}\otimes O_{12} \cr 	
{\rm with} ~ (V'_{\rm CKM})_{13} = 0 \cr}$
& $\surd$ & $\surd$ & & & $\surd$ & $\surd$ \\ \hline
$\matrix{V'_{\rm CKM} = O_{12}\otimes O_{23} \cr
{\rm with} ~ (V'_{\rm CKM})_{31} = 0 \cr}$ 	
& & $\surd$ & $\surd$ & $\surd$ & $\surd$ & \\ \hline
$\matrix{V'_{\rm CKM} = O_{13}\otimes O_{12} \cr
{\rm with} ~ (V'_{\rm CKM})_{23} = 0 \cr}$ 	
& $\surd$ & & $\surd$ & & $\surd$ & $\surd$ \\ \hline
$\matrix{V'_{\rm CKM} = O_{12}\otimes O_{13} \cr
{\rm with} ~ (V'_{\rm CKM})_{32} = 0 \cr}$ 	
& & $\surd$ & $\surd$ & $\surd$ & & $\surd$ 
\end{tabular}
\end{center}
\end{table}

\newpage

\begin{table}
\caption{The impact of quark mass degeneracy on the CKM matrix. The
symbol ``$\surd$'' means that it is possible to make two of the
three mixing angles of $V^{~}_{\rm CKM}$ vanishing in the limit where 
a pair of up-type quarks and a pair of down-type quarks are respectively
degenerate in mass.}
\vspace{0.3cm}
\begin{center}
\begin{tabular}{|c|c|c|c|}
& $V''_{\rm CKM} = O_{12}$ & $V''_{\rm CKM} = O_{23}$ 
& $V''_{\rm CKM} = O_{13}$ \\ \hline
~~ $m_u = m_c$ $\&$ $m_d = m_s$ ~~ &
& $\surd$ & $\surd$ \\ \hline
$m_u = m_c$ $\&$ $m_d = m_b$ &
& $\surd$ & \\ \hline
$m_u = m_c$ $\&$ $m_s = m_b$ &
& & $\surd$ \\ \hline
$m_u = m_t$ $\&$ $m_d = m_s$ &
& $\surd$ & \\ \hline
$m_u = m_t$ $\&$ $m_d = m_b$ &
$\surd$ & $\surd$ & \\ \hline
$m_u = m_t$ $\&$ $m_s = m_b$ &
$\surd$ & & \\ \hline
$m_c = m_t$ $\&$ $m_d = m_s$ &
& & $\surd$ \\ \hline
$m_c = m_t$ $\&$ $m_d = m_b$ &
$\surd$ & & \\ \hline
$m_c = m_t$ $\&$ $m_s = m_b$ &
$\surd$ & & $\surd$ 
\end{tabular}
\end{center}
\end{table}

\newpage

\begin{table}
\caption{The impact of lepton mass degeneracy on the MNS matrix. The
symbol ``$\surd$'' means that it is possible to make one off-diagonal
matrix element of $V^{~}_{\rm MNS}$ vanishing in the limit where two 
charged leptons are degenerate in mass or two neutrinos have the
same mass and the same Majorana phase. Note that the complex neutrino
masses $m'_a \equiv m_a e^{2i\phi_a}$ (for $a=1,2,3$) contain the
information about Majorana-type CP-violating phases.}
\vspace{0.3cm}
\begin{center}
\begin{tabular}{|c|c|c|c|c|c|c|}
& $m_e = m_\mu$ & $m_e = m_\tau$ & $m_\mu = m_\tau$
& $m'_1 = m'_2$ & $m'_1 = m'_3$ & $m'_2 = m'_3$ \\ \hline
$\matrix{U'_{\rm MNS} = O_{23}\otimes O_{13} \cr
{\rm with} ~ (U'_{\rm MNS})_{12} = 0 \cr}$
& $\surd$ & $\surd$ & & & & \\ \hline
$\matrix{U'_{\rm MNS} = O_{13}\otimes O_{23} \cr
{\rm with} ~ (U'_{\rm MNS})_{21} = 0 \cr}$
& $\surd$ & & $\surd$ & & & \\ \hline
$\matrix{U'_{\rm MNS} = O_{23}\otimes O_{12} \cr 	
{\rm with} ~ (U'_{\rm MNS})_{13} = 0 \cr}$
& $\surd$ & $\surd$ & & & & \\ \hline
$\matrix{U'_{\rm MNS} = O_{12}\otimes O_{23} \cr
{\rm with} ~ (U'_{\rm MNS})_{31} = 0 \cr}$ 	
& & $\surd$ & $\surd$ & & & \\ \hline
$\matrix{U'_{\rm MNS} = O_{13}\otimes O_{12} \cr
{\rm with} ~ (U'_{\rm MNS})_{23} = 0 \cr}$ 	
& $\surd$ & & $\surd$ & & & \\ \hline
$\matrix{U'_{\rm MNS} = O_{12}\otimes O_{13} \cr
{\rm with} ~ (U'_{\rm MNS})_{32} = 0 \cr}$ 	
& & $\surd$ & $\surd$ & & & \\ \hline\hline
~ $\matrix{U'_{\rm MNS} = O_{23}\otimes O_{13} \otimes P_2 \cr
{\rm with} ~ (U'_{\rm MNS})_{12} = 0 \cr}$ ~
& & & & $\surd$ & & $\surd$ \\ \hline
$\matrix{U'_{\rm MNS} = O_{13}\otimes O_{23} \otimes P_1 \cr
{\rm with} ~ (U'_{\rm MNS})_{21} = 0 \cr}$
& & & & $\surd$ & $\surd$ & \\ \hline
$\matrix{U'_{\rm MNS} = O_{23}\otimes O_{12} \otimes P_3 \cr 	
{\rm with} ~ (U'_{\rm MNS})_{13} = 0 \cr}$
& & & & & $\surd$ & $\surd$ \\ \hline
$\matrix{U'_{\rm MNS} = O_{12}\otimes O_{23} \otimes P_1 \cr
{\rm with} ~ (U'_{\rm MNS})_{31} = 0 \cr}$ 	
& & & & $\surd$ & $\surd$ & \\ \hline
$\matrix{U'_{\rm MNS} = O_{13}\otimes O_{12} \otimes P_3 \cr
{\rm with} ~ (U'_{\rm MNS})_{23} = 0 \cr}$ 	
& & & & & $\surd$ & $\surd$ \\ \hline
$\matrix{U'_{\rm MNS} = O_{12}\otimes O_{13} \otimes P_2 \cr
{\rm with} ~ (U'_{\rm MNS})_{32} = 0 \cr}$ 	
& & & & $\surd$ & & $\surd$ 
\end{tabular}
\end{center}
\end{table}

\newpage

\begin{table}
\caption{The impact of quark mass degeneracy on the MNS matrix. The
symbol ``$\surd$'' means that it is possible to make two of the
three mixing angles of $V^{~}_{\rm MNS}$ vanishing in the limit where 
two charged leptons are degenerate in mass and two neutrinos have the
same mass and the same Majorana phase. Note that the complex neutrino
masses $m'_a \equiv m_a e^{2i\phi_a}$ (for $a=1,2,3$) contain the
information about Majorana-type CP-violating phases.}
\vspace{0.3cm}
\begin{center}
\begin{tabular}{|c|c|c|c|}
& $U''_{\rm MNS} = O_{12}$ & $U''_{\rm MNS} =O_{23}$ & 
$U''_{\rm MNS} = O_{13}$ \\ \hline
~~ $m_e = m_\mu$ $\&$ $m_1' = m_2'$ ~~ &
& $\surd$ & $\surd$ \\ \hline
$m_e = m_\mu$ $\&$ $m_1' = m_3'$ &
& $\surd$ & \\ \hline
$m_e = m_\mu$ $\&$ $m_2' = m_3'$ &
& & $\surd$ \\ \hline
$m_e = m_\tau$ $\&$ $m_1' = m_2'$ &
& $\surd$ & \\ \hline
$m_e = m_\tau$ $\&$ $m_1' = m_3'$ &
$\surd$ & $\surd$ & \\ \hline
$m_e = m_\tau$ $\&$ $m_2' = m_3'$ &
$\surd$ & & \\ \hline
$m_\mu = m_\tau$ $\&$ $m_1' = m_2'$ &
& & $\surd$ \\ \hline
$m_\mu = m_\tau$ $\&$ $m_1' = m_3'$ &
$\surd$ & & \\ \hline
$m_\mu = m_\tau$ $\&$ $m_2' = m_3'$ &
$\surd$ & & $\surd$ 
\end{tabular}
\end{center}
\end{table}
\end{document}